# The Minimum Bandwidth of Narrowband Spikes in Solar Flare Decimetric Radio Waves

Peter Messmer, Arnold O. Benz

Institute of Astronomy, ETH, CH-8092 Zürich, Switzerland

**Abstract.** The minimum and the mean bandwidth of individual narrowband spikes in two events in decimetric radio waves is determined by means of multi-resolution analysis. Spikes of a few tens of millisecond duration occur at decimetric/microwave wavelength in the particle acceleration phase of solar flares. A first method determines the dominant spike bandwidth scale based on their scalegram, the mean squared wavelet coefficient at each frequency scale. This allows to measure the scale bandwidth independently of heuristic spike selection criteria, e.g. manual selection. The major drawback is a low resolution in the bandwidth. To overcome this uncertainty, a feature detection algorithm and a criterion for spike shape in the time-frequency plane is applied to locate the spikes. In that case, the bandwidth is measured by fitting an assumed spike profile into the denoised data. The smallest FWHM bandwidth of spikes was found at 0.17% and 0.41% of the center frequency in the two events. Knowing the shortest relevant bandwidth of spikes, the slope of the Fourier power spectrum of this two events was determined and no resemblance to a Kolmogorov spectrum detected. Additionally the correlation between spike peak flux and bandwidth was examined.

## 1. Introduction

Narrowband spikes in decimeter radio waves (a review on observations can be found in Benz 1986) are closely associated with hard X-ray emission and thus with energy release of impulsive solar flares (Benz & Kane 1986; Güdel et al. 1991; Aschwanden & Güdel 1992). Only a few percent of the HXR flares are associated with spikes. If they occur, they appear in large numbers (several thousand) and are scattered over a frequency band $\nu_{\min}/\nu_{\max} \lesssim 2$. The most salient feature of spikes is their small bandwidth of a few percent. The bandwidth of a single spike is given by (i) the intrinsic bandwidth of the emission process times (ii) the variation of the emission frequency over the source region, caused by the range of some plasma parameters. The central frequency is determined by the relevant characteristic frequency such as the plasma frequency $\omega_p$ or the electron gyro-frequency $\Omega_e$. Thus, the second factor limits the source size for a given scale length of the relevant plasma parameter (density or magnetic field). The first factor severely limits the possible emission mechanisms. Broadband radiations, such as gyro-synchrotron emission, are excluded. Plasma emission of Langmuir waves is questionable. Most favored in the past has been the electron maser mechanism (Holman et al. 1980; Melrose & Dulk 1982; Fleishman 1994). It has an intrinsic width of

$$\frac{\Delta \nu}{\nu} \approx 0.5 (\frac{v_{\max}}{c})^2 \Delta \alpha \sin(2\alpha_c) \approx (\frac{v_{\text{hot}}}{c})^2 \quad (1)$$

(Hewitt & Melrose 1985), where $v_{\max}$ is the center of the resonance circle in velocity space of the fastest growing waves, $\alpha_c$ is the loss-cone angle, $\Delta \alpha$ the angle (in radians), over which the velocity distribution in the loss-cone falls off, and $v_{\text{hot}}$ the characteristic velocity of the energetic electrons.

However, it is not clear how electron maser emission can originate in the solar corona, where generally $\omega_p/\Omega_e > 1$, and how it can escape from there in view of the gyro-resonance absorption on thermal electrons. As an alternative, emission by the interaction of Bernstein modes and upper-hybrid waves has been proposed (Willes & Robinson 1996). Nevertheless, the emission process of spikes remains unclear.

The minimum bandwidth sets the most severe requirements on the emission process and limits the source size. Bandwidths of individual bursts have been measured by Benz (1986) and more elaborately by Csillaghy & Benz (1993). The latter authors have fitted Gaussians to single spikes and reported a smallest bandwidth of 3.5 MHz FWHM at 350 MHz center frequency (event of 82/06/04) and of 2.5 MHz FWHM at 800 MHz (83/06/06). The reported mean values are 7.32 and 7.04 MHz, respectively.

Here measurements of the minimum and the mean bandwidth of spikes by statistical methods are reported. Two different methods were applied, both based on the wavelet representation of the measured spectra.

In Section 2, the wavelet representation of a sampled signal is introduced. Special properties of the scalegram, an analogon to the Fourier power spectrum in case of wavelet representation, are described in Sect. 3 and the method is applied to test data. In Section 4, the analyzed data set is introduced. The spike bandwidth scale is then determined, based on the scalegram of the individual spectra in Sect. 5.

In Sect. 6, the wavelet representation is used to denoise the individual spectra. An assumed spike profile is fitted into the smoothed signal, yielding a spike bandwidth distribution, including the smallest significant spike bandwidth.

Both results are compared to Fourier transforms in Section 7, and the implications discussed in Sections 8 and 9. Conclusions are given in Section 10.

## 2. Multi-resolution Analysis Algorithm

In this section, a rough description of the multi-resolution analysis algorithm is presented. For more mathematical details, see Mallat (1989), Daubechies (1988) or Cohen (1992).

The basic idea of multi-resolution representation of a sampled signal is to decompose it into a smoothed signal at lower resolution and a signal containing the details removed by the smoothing process. Iteration of this process on the smoothed signal leads finally to a smooth signal and a set of details at different resolutions. This representation is useful to investigate signal structures at different scales.

The multi-resolution analysis is governed by the so called *smoothing function* $q(x)$ which determines, how the signal sampled at intervals $\Delta x$ is converted into one at lower resolution, sampled at intervals $2^j \Delta x$, where the exponent $j$ stands for the resolution scale. If the smoothing function satisfies certain conditions (Mallat 1989), then the detail signal can be expressed in a wavelet basis.

A wavelet basis is derived from a so called *mother wavelet* $w(x)$ by scaling and translation, according to

$$w_k^j(x) = 2^{-j/2} w\left(\frac{x - b(k,j)}{a(k,j)}\right) = 2^{-j/2} w\left(\frac{x - 2^j \Delta x \cdot k}{2^j}\right) \quad (2)$$

with the common scaling factor $a = 2^j$ and translation $b = 2^j \Delta x \cdot k, k \in \mathbb{Z}$. A signal $f(x)$, consisting of $N$ samples at interval $\Delta x$, is expressed in this wavelet basis by

$$f(x) = \sum_j \sum_k d_k^j w_k^j(x) \quad j, k \in \mathbb{Z} \quad (3)$$

with the wavelet coefficients

$$d_k^j = \int f(x) w_k^j(x) dx. \quad (4)$$

The detail signal at resolution $j$, which was removed from the smoothed signal at resolution $j+1$, can be expressed as a superposition of the wavelets $w_k^j(x)$ for given $j$ and $k \in \mathbb{Z}$.

By iteration of this process on the smoothed signal, the original signal can be decomposed in the following way

$$f(x) = \sum_k c_k^{j_{\max}} q_k^j(x) + \sum_{j=0}^{j_{\max}} \sum_k d_k^j w_k^j(x) \quad (5)$$

for a certain $j_{\max} \leq \log_2 N$, which determines the coarsest resolution. The computation of the wavelet coefficients can be performed in an iterative way, making the evaluation of the integrals Eq. 4 necessary at the finest resolution only.

Obviously, the mother wavelet depends on the smoothing function, but there exist systematic approaches to determine the mother wavelet for a given smoothing function. In this paper, a smoothing function and their corresponding wavelet was applied, which has already been used by Bendjoya et al. (1993), Schwarz et al. (1998) and Aschwanden et al. (1998). This wavelet, sometimes called triangle wavelet, is very compact and allows to analyze signals down to their shortest scales. The set of smoothing functions and wavelets is given by

$$w(x) = \begin{cases} \frac{1}{2} - \frac{3|x|}{4\Delta x} & \text{for} \quad \frac{|x|}{\Delta x} \leq 1, \\ \frac{|x|}{4\Delta x} - \frac{1}{2} & \text{for} \quad 1 < \frac{|x|}{\Delta x} \leq 2, \\ 0 & \text{otherwise} \end{cases} \quad (6)$$

$$q(x) = \begin{cases} \frac{1}{2} - \frac{|x|}{2\Delta x} & \text{for} \quad \frac{|x|}{\Delta x} \leq 1, \\ 0 & \text{otherwise} \end{cases} \quad (7)$$

$$w_k^j(x) = w[2^{-(j-1)}(x - k\Delta x)], \quad j = 1, 2, \ldots, k \in Z \quad (8)$$

$$q_k^0 = 2q(x - k\Delta x), \quad (9)$$

$$q_k^j(x) = q[2^{-j}(x - k\Delta x)], \quad j = 1, 2, \ldots, k \in Z. \quad (10)$$

The smoothing function coefficients at the finest resolution can be computed by

$$c_k^0 = \int f(x) q_k^0(x) dx = \sum_i f(i\Delta x) q_k^0(i\Delta x), \quad (11)$$

what means that the $c_k^0$ are directly the sampled data values, as can be seen by inserting the definition of $q_k^0$. At all coarser resolutions $j > 0$, the smoothing function coefficients are given by the recursion

$$c_k^j = \frac{1}{2} c_k^{j-1} + \frac{1}{4} \left( c_{k-2^{j-1}}^{j-1} + c_{k+2^{j-1}}^{j-1} \right) \quad (12)$$

By inserting the definition of the wavelet coefficients, it becomes apparent that

$$d_k^j = \sum_i c_i^{j-1} w_k^j(x) = c_k^{j-1} - c_k^j \quad (13)$$

The iteration stops at $j_{\max} < \log_2 N$.

In the following, the wavelet representation is used in several ways: both on signals in frequency and time direction, and both for feature detection and as a smoothing tool.

## 3. Scalegram properties

Wavelet coefficients $d_k^j$ are usually displayed in a $2^j - k$ plane as so called scalograms (Schwarz et al. 1998). If only spectral information is relevant, the squared wavelet coefficients are plotted as the $k$-average for all $j$, the so called scalegram $s(j)$ (Scargle et al. 1993),

$$s(j) = \frac{2^j}{N} \sum_{k}^{N-2^j} \left( d_{2^j k}^j \right)^2. \quad (14)$$

Structures in the scale range $[2^{j-1}, 2^j]$ contribute mainly to the scalegram value $s(j)$ at frequency scale $2^j$.

First it should be noted that adding a constant offset to the analyzed spectrum has no influence on its scalegram (Kliem et al. 1998). On the other hand, scaling the spectrum by a constant factor leads to a shift of all scalegram values, but does not change the overall scalegram shape.

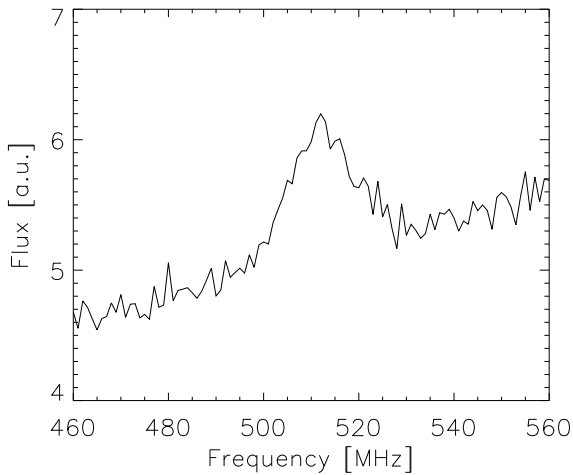

**Fig. 1.** An example of an artificial spectrum including the superposition of noise, ramp and a Gaussian with 8 MHz FWHM.

To give an overview of scalegrams, scalegrams of different simulated spectra are compared: Pure noise, linear ramp, superposition of both and superpositions of noise, ramp and Gaussians with different widths.

The noise spectra $x_n(\nu)$ are given by normally distributed random numbers with standard deviation 0.09, the linear ramp is given by $x_r(\nu) = 0.01 \cdot \nu$ and the Gaussians $x_g^m(\nu) = \exp(-4 \ln 2 \cdot \nu^2 / m^2)$ with $m \in \{4, 8, 16\}$ MHz FWHM, with amplitude 1, resulting in a signal-to-noise ratio of 11.1. The total width is always 1024 MHz and the sampling interval 1 MHz. Due to the random nature of the noise spectrum, all scalegrams are computed for 10 different noise spectra and averaged. Fig. 1 displays a part of such a simulated spectrum.

The scalegrams of the different spectra are compared in Fig. 2: The scalegram of pure noise, $x_n$, shows most weight at smallest scales, whereas the linear ramp $x_r$ shows most weight at largest scales. The simulated background spectrum $x_n + x_r$ yields a scalegram with large weights at both smallest and largest scales and less weight at medium scales. In case of the Gaussians, the error bars indicate the 99% confidence interval for the mean scalegram values. The error bars for the noisy spectra indicate the 3-fold standard deviation ($3\sigma$) of the $s(j)$ distribution determined from the 10 trials.

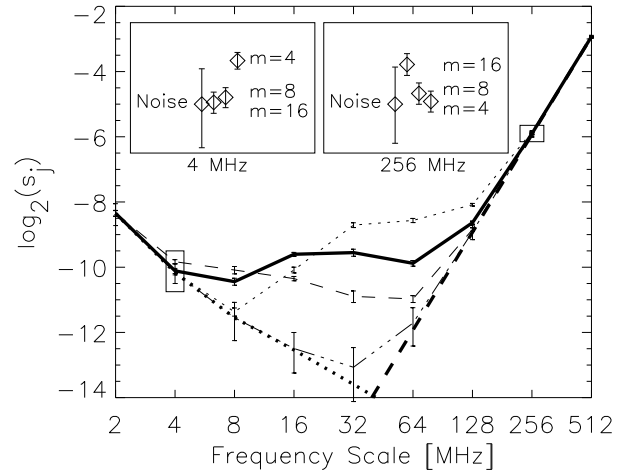

**Fig. 2.** The scalegrams of artificial spectra for noise (dotted, thick), linear ramp (dashed, thick), their superposition (dashed-dotted) and for the superposition of noise, ramp and a Gaussian with width $m = 4$ MHz (dashed), 8 MHz (solid, thick), and 16 MHz (dotted) FWHM. The logarithm to basis 2 of the mean squared wavelet coefficient is displayed versus the logarithmic frequency scale in $2^j$ MHz. The insets show zooms of the scalegram at scales 4 MHz (left) and 256 MHz (right). The data points in the insets are spread horizontally for better viewing. The error bars are discussed in the text.

The mean scalegram of $x_g^{16} + x_r + x_n$ is within $3\sigma$ of the noise distribution for scales up to 8 MHz (cf. Fig. 2, label $m = 16$). For larger scales, it differs significantly from the simulated background scalegram up to 256 MHz (Fig. 2, right inset).

The case of $x_g^4 + x_r + x_n$ is considerably different from the noise scalegram already at scales of 4 MHz (see Fig.2, left inset), and merges with the scalegram of the background at 128 MHz.

The case for $x_g^8 + x_r + x_n$ shows the smooth transition between the two other cases.

These examples demonstrate how the scalegram can be used to estimate the scale of a dominant structure in a noisy signal by detecting significant deviation (e.g. $3\sigma$) from the mean noise scalegram at smallest frequency scales.

However this property may lead to the assumption that the FWHM of a single Gaussian can be determined by choosing the smallest significant scale exceeding the background noise scalegram. Unfortunately, this is not true: Assuming a single Gaussian in a spectrum overlayed by noise, the smallest significant scale depends also on the signal-to-noise ratio of the Gaussian compared to the background noise. Fig. 3 shows the minimal signal-to-noise ratio which is necessary to make a single Gaussian with a given FWHM significant at a given scale. For every width and for every amplitude of the Gaussian, 80 noise spectra were computed and averaged.

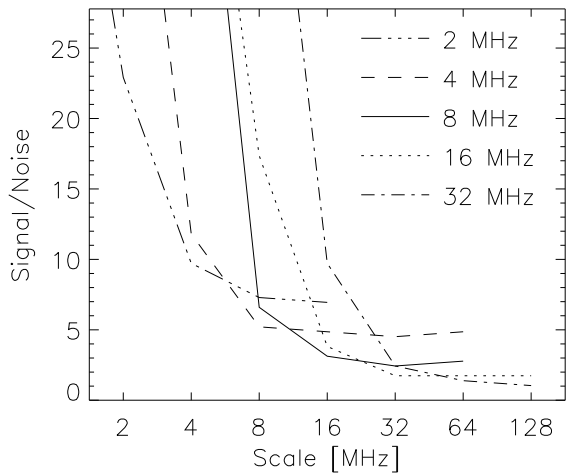

**Fig. 3.** Minimal amplitude of a single Gaussian to turn its scalegram value significant at a given scale, compared to the background noise.

A single Gaussian with a width of 2 MHz does not become significantly different from the background noise at any scale unless its signal to noise ratio exceeds 6. The weakest cases become only significantly different from the background at scale 16 MHz. On the other hand, in case of strong signals, the smallest significant scale is 2 MHz.

Assuming a single Gaussian in a spectrum, a significant scalegram value at scale $2^j$ can have different origins: either a relatively weak Gaussian with width $< 2^j$ or a Gaussian with larger amplitude of width $2^j$ or it can even originate from a Gaussian with width $> 2^j$ with yet larger amplitudes.

For all these reasons, a direct interpretation of the smallest significant scale as FWHM bandwidth of a single Gaussian is not possible.

Nevertheless, for bandwidths $< 32$ MHz and for a signal-to-noise ratio up to 26, typical for the data under consideration, the smallest significant scale can be interpreted as an upper limit of the underlying spectral structure.

## 4. Observational Data

The data was recorded with the frequency agile radio spectrometer IKARUS (Perrenoud 1982) of ETH Zürich at Bleien Observatory. The analyzed data set consists of two events, one recorded on 83/06/06, after 13:59:14.0 UT, between 873-1000 MHz (event 1) and one recorded on 82/06/04 at 13:38:30.0 UT between 320-383 MHz (event 2). The spectral resolution in both cases was 1 MHz. Both events have been classified in the spectrogram as decimetric narrowband spikes.

Light-curves of a representative frequency channel are given in Fig. 4 for illustration.

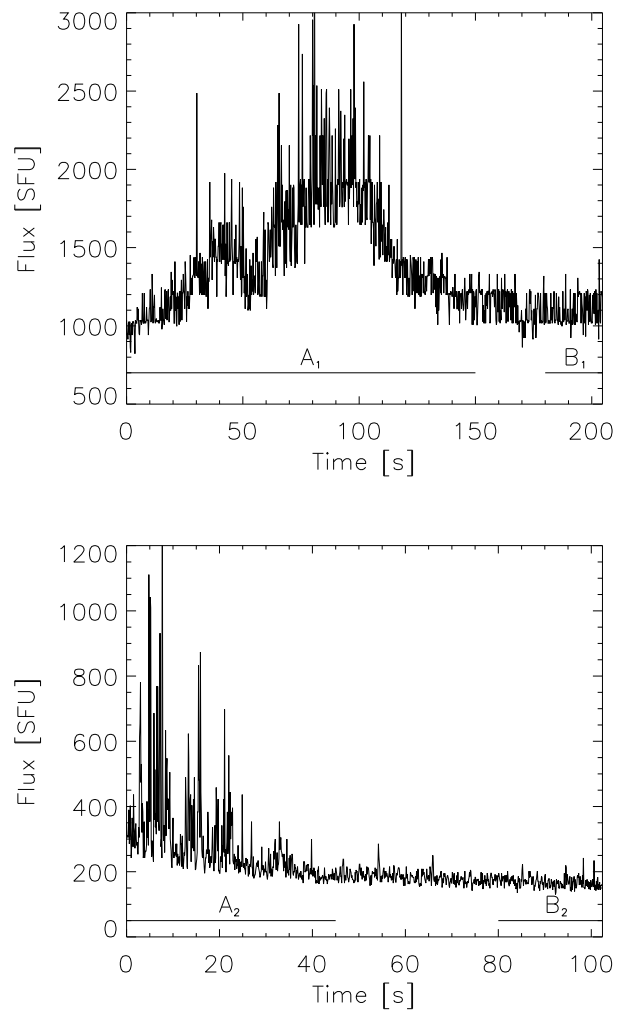

**Fig. 4.** Lightcurves of the analyzed spike events. Top: event 1, recorded on 83/06/06 at 990 MHz after 13:49:14.0 UT with a temporal resolution of 0.2s. Bottom: event 2, recorded on 82/06/04 after 13:38:30.0 UT with a temporal resolution of 0.1s. The intervals labeled $A_i$ and $B_i$ refer to the analyzed burst and background data.

In both cases two intervals are investigated: $A_i$, which contains mainly short, intense structures on an enhanced background, and $B_i$ which contains mainly background noise.

The interval $A_1$ in event 1 contains 750 spectra and $B_1$ 123, each of them containing 128 frequency channels. Event 2 consists of 450 spectra in the interval $A_2$ and 223 in interval $B_2$ with 64 frequency channels each.

## 5. Scalegrams of Decimetric Spikes

The aim is to distinguish spikes from noise in both events introduced in the previous section and to determine the minimal and the dominant spike bandwidth scale. One major problem is that in both cases the spikes are superposed on a background radiation which varies in time by up to a factor 2 (see Fig. 4). This leads to an enhancement and variation in noise amplitude, what in turn has an influence on the resulting scalegram. If scalegrams of spectra at different times are compared, this effect has to be taken into account. This problem is considered first.

For this purpose, the smoothing properties of MRA were used in *time domain*, splitting every light-curve into a smooth signal $f_\nu(t)$ and a detail signal $\Delta f_\nu(t)$

$$F_\nu(t) = f_\nu(t) + \Delta f_\nu(t). \quad (15)$$

The decomposition was made on a scale large compared to a mean spike duration. In both cases it was at $j^{\max} = 3$, corresponding to structures of 1.6s in event 1 and 0.8s in event 2.

Every frequency channel was decomposed in this fashion and the normalized noise

$$g_t(\nu) = \frac{\Delta f_\nu(t)}{f_\nu(t)} \quad (16)$$

was determined. Figure 5 shows examples of the resulting spectra $g_t(\nu)$ for both events. The normalized signal-to-normalized noise ratio of the spectra is < 17 in event 1 and < 26 in event 2. These spectra $g_t(\nu)$ were then analyzed by means of MRA in *frequency domain*.

In a first step, the spectral scalegrams of the noise intervals $B_i$ are computed and averaged for every frequency scale (see Fig. 6). In a second step, the scalegrams in the intervals $A_i$ were computed. Each scalegram was assigned to one of two categories: if one of the scalegram values exceeded the $3\sigma$ level above the mean noise scalegram, it was considered as containing a spike. If none of the scalegram values exceeds $3\sigma$, it was considered as containing no spike.

In the following, the scalegrams of spectra with spikes are compared to those containing background noise. Both Figures 6 top and bottom show good agreement of the mean noise scalegram with the mean scalegram of spectra containing no spike. For small scales up to 8 MHz, the agreement is nearly perfect, but also at larger scales,

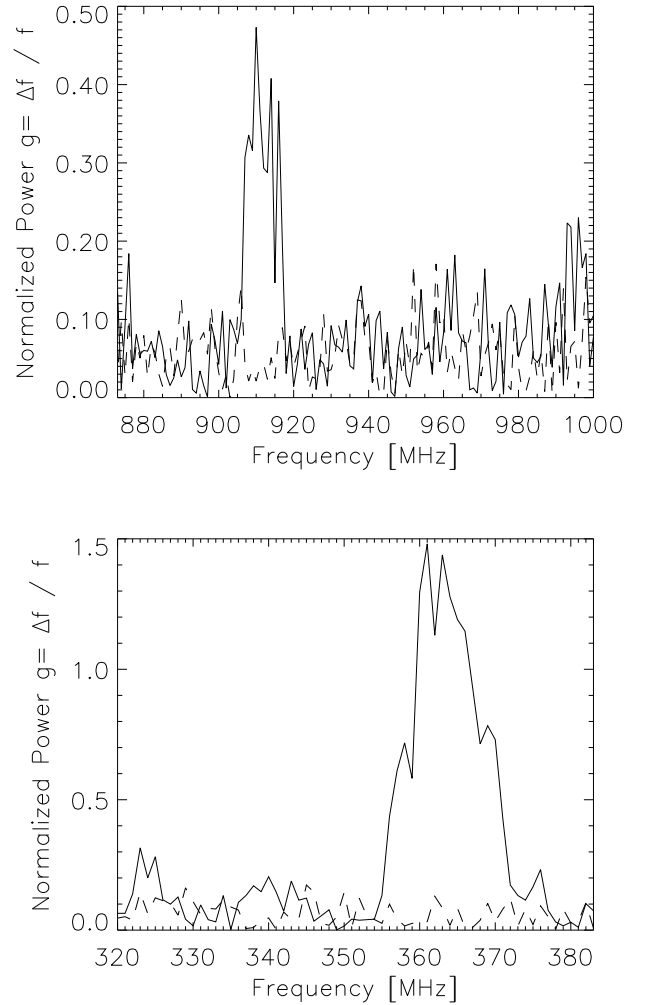

**Fig. 5.** Examples of spectra for both event 1 (top) and event 2 (bottom). Both a spectrum from interval $A_i$, containing a spike (solid), and one from interval $B_i$, containing no spike (dashed) are presented.

the 99% confidence intervals for both populations overlap almost everywhere.

In event 1 (Fig. 6 top) at the shortest scale (2 MHz), the mean scalegram of spectra containing spikes coincides with the mean scalegram of noise. This indicates no difference at the shortest scale between spectra containing spikes and those not containing spikes. However, 10 spectra with spikes exceeding $3\sigma$ of the background distribution at scale 2 MHz were found. At scale 8 MHz, the mean spike-containing scalegram exceeds the mean noise scalegram by $3\sigma$, indicating the average smallest significant scale for spikes to be 8 MHz.

In the bottom panel of Fig. 6, the spike-containing scalegrams differ significantly from those not containing a spike down scale 8 MHz. It is marginally significant already at 4 MHz. At the shortest scale, the confidence intervals of both mean values overlap, indicating no significant scalegrams of the spectra, but the scalograms, the $2^j$-$k$ plots of each spectrum's wavelet coefficients, are analyzed. In these scalograms, a feature detection algorithm was applied, comparable to the one implemented by Bendjoya et al. (1993): In a first step, the mean scalogram value for every frequency scale in the intervals $B_i$ were computed. Next, the scalogram for every spectrum in the interval $A_i$ was computed. The scalogram was then searched for *seeds*, points exceeding the mean scalogram value by $5\sigma$. To avoid reconstruction biases, up to $2^{j-1}$ neighboring wavelet coefficients exceeding the mean scalogram value were added on both sides of the seeds. From this scalogram, a smoothed spectrum was reconstructed. After this denoising procedure, a Gaussian was fitted into every single spike profile. The distribution of the FWHM of these Gaussians is shown in Fig.7.

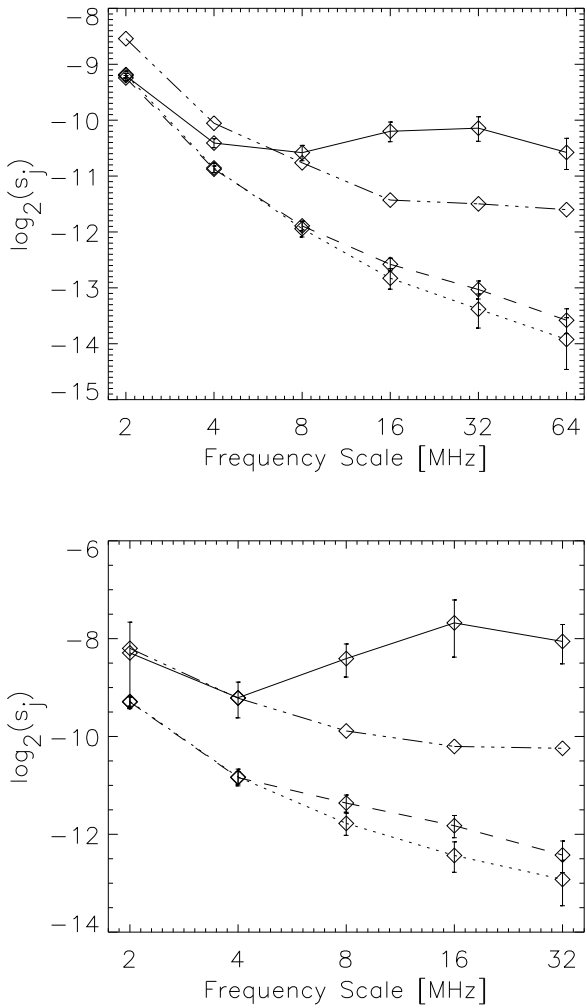

**Fig. 6.** Scalegrams of event 1 (top) and event 2 (bottom). Dotted: Mean scalegram of interval $B_i$. Dashed: The mean scalegram of spectra at interval $A_i$ with no spikes. Solid: Mean spike scalegram. Dash-dotted curve: 3 standard deviations above the mean scalegram of interval $B_i$. The error-bars indicate the 99% confidence interval of the mean values.

difference from the noise distribution. Again several spectra exceeding the background distribution by more than $3\sigma$ were found.

The 8 MHz mean structure size of spikes agrees well with the results reported in (1993). However, the smallest structure scale exceeding noise determined here is 2 MHz in both events. The earlier reported lower bound was 3.5 MHz, resp. 2.5 MHz. This may be due to the manual selection favoring larger scales.

## 6. Individual Spike Analysis

Contrary to the previous section, where statistical properties of spike scalegrams were investigated, the bandwidth of individual spikes is now determined. Therefore not the

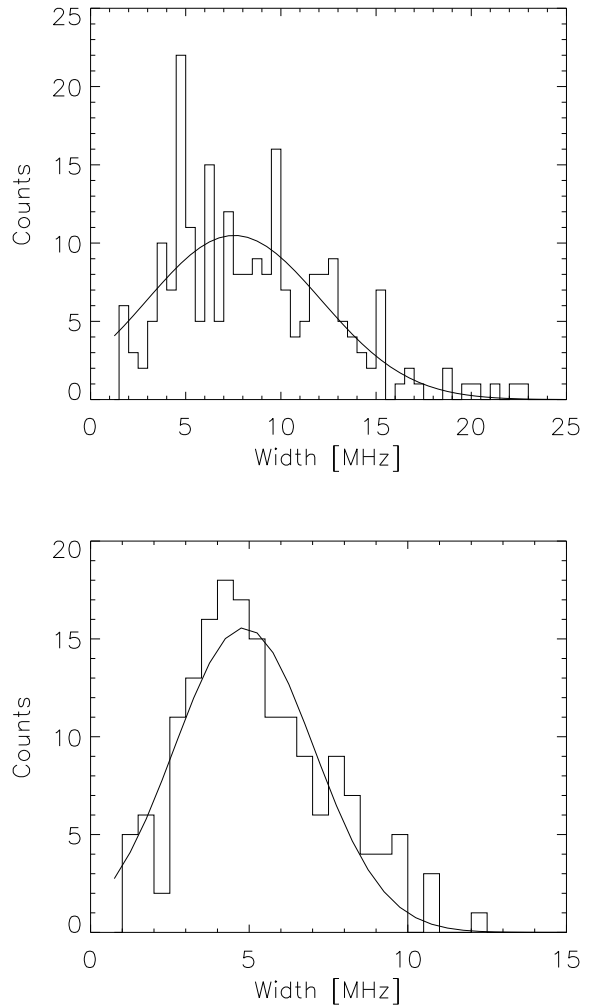

**Fig. 7.** Number of spikes per bandwidth for individual spikes in event 1 (top) and event 2 (bottom). The solid line represents a Gaussian, fitted to the distribution.

|  |  | Event 1 | Event 2 |
|---|---|---|---|
| Date |  | 83/06/06 | 82/06/04 |
| Time | UT | 13:59:14.0 | 13:38:30.0 |
| Number of sweeps | $A_i$ | 750 | 450 |
|  | $B_i$ | 123 | 223 |
| Frequency range | [MHz] | 873 - 1000 | 320 - 383 |
| Number of spikes |  | 230 | 174 |
| Mean scale (MRA) | [MHz] | 8 | (4) 8 |
| Mean FWHM | [MHz] | 7.5 | 4.8 |
| Minimal scale (MRA) | [MHz] | 2 | 2 |
| Minimal FWHM | [MHz] | 1.5 | 1.2 |

**Table 1.** Summary of the event parameters. The areas $A_i$ contain the burst spectra, $B_i$ the background spectra. The mean and minimal spike bandwidth scale was determined by multiresolution analysis, whereas the mean and minimal FWHM was determined by fitting Gaussians into the denoised spectra.

Event 1 consisted of 230 individual spikes, which lay entirely in the observed frequency band. They have a peak in the bandwidth distribution at 7.5 MHz and a distribution width of 4.6 MHz. The minimal detectable bandwidth was 1.5 MHz. Event 2 consisted of 174 individual spikes entirely in the observed frequency band, with a peak in the bandwidth distribution at 4.8 MHz and a distribution width of 2.2 MHz. The minimal detectable bandwidth was 1.2 MHz.

These values agree well with the bandwidth scales determined by the scalegram analysis: The mean spike has a width of about 8 MHz in event 1, and 4 MHz in event 2 respectively.

## 7. Power Spectrum Analysis

Fourier power spectra of spikes in frequency have been reported to obey a power law (Karlicky et al. 1996). To determine the functional form, it is essential to know the relevant range of the Fourier spectrum.

Knowing the range of frequency scales of spikes, upper and lower limits of the relevant Fourier range can be determined and in turn the power-law exponent can be computed.

For every frequency scan containing spikes $g_\nu$ its Fourier transform and the resulting power spectrum was computed. All spectra in the intervals $A_i$ and $B_i$ were then averaged, resulting in the mean power spectra plotted in Fig. 8.

In the previous sections, the scales for emitting structures were measured to be in the ranges 8-64 MHz and 4-32 MHz for the two events respectively. This corresponds to $-1.81 < \log k_f < -0.90$, $-1.51 < \log k_f < -0.60$ respectively, where $k_f$ [MHz$^{-1}$] is the Fourier frequency.

In event 1 the spike power spectrum coincides with noise at high frequencies, at about the upper limit of the significant range. This agrees with the insignificance of the scalegram values at small frequency scales in Fig. 6 (top). The spike scalegram for event 2 in Fig. 6 (bottom) does not show such a good overlap at smallest scales. This is also reflected in Fig. 8 (bottom), where however the confidence intervals of background noise and spike seem to overlap.

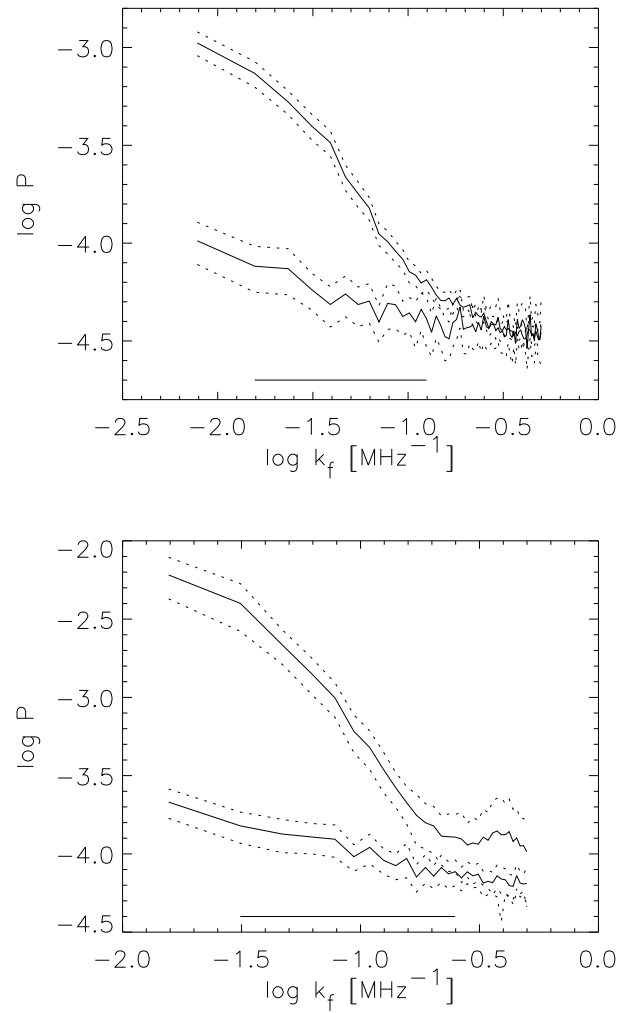

**Fig. 8.** Power spectra of both event 1 (top) and event 2 (bottom). Upper solid curve: Mean power spectra during interval $A_i$. Lower solid curve: Mean noise spectra during interval $B_i$. Dotted curves: 99% confidence interval. Horizontal line: Significant range of spike spectrum, derived from the scalegrams.

## 8. Scale Height

Assuming the electron density to decrease exponentially with altitude, and assuming fundamental or harmonic plasma frequency as the spike emission frequency, the frequency scale of radio spectra can be transformed into altitude $h$ in the solar atmosphere by

$$h(f) = -2H \ln \frac{f}{f_0} \qquad (17)$$

where $H$ is the density scale height, $f$ the emission frequency, $h$ the altitude and $f_0$ the emission frequency at $h = 0$.

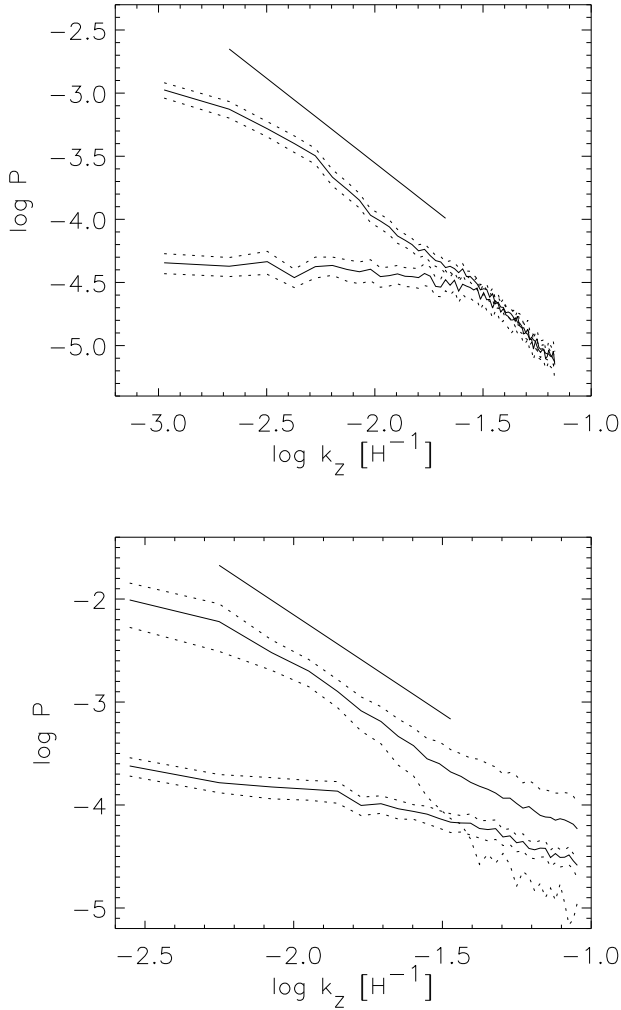

Fig. 9. Transformation of spectra in the frequency domain into spectra in the spatial domain under assumption of exponential decrease in electron density for event 1 (top) and event 2 (bottom). The curve labeling is identical to that of Fig. 8, the straight line indicates the power-law fit, excluding the smallest frequencies $k_z$. The units of $k_z$ are the inverse of the density scale height H.

The interpolated spectra are resampled according to the new scale and again the mean power spectrum is computed for both cases. Fig. 9 shows the spectra, both of them exhibiting power-law characteristics. In the significant range, their slope is $\log P/\log k_z = -1.34 \pm 0.03$ and $-1.92 \pm 0.04$ for event 1 and 2, respectively. This fit does not include the smallest frequencies $k_z$, which are likely influenced by the boundaries. Including them blindly, the resulting power-law exponent would be $\log P/\log k_z = -1.21 \pm 0.05$ and $-1.67 \pm 0.09$. It shows how sensitive the exponents of a fitted power law react on the range of considered frequencies.

## 9. Spike Peak Flux

Each individual spectrum of spikes was searched for the smallest significant scale $2^j$ MHz by MRA, as described in Sect. 5. Then the average peak flux $A_j$ of all spectra with smallest significant scale $2^j$ MHz was determined. Fig. 10 compares the inverse average peak fluxes $1/A_j$ to the smallest significant frequency scale $2^j$ MHz.

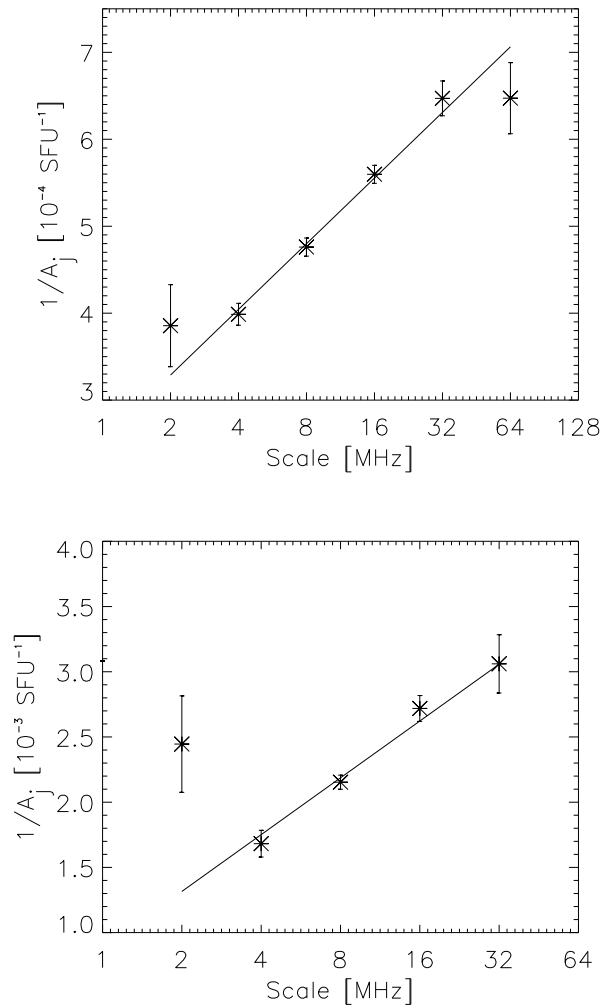

Fig. 10. Comparison of smallest significant scale and inverse mean peak flux $1/A_j$ for both event 1 (top) and event 2 (bottom) determined by MRA.

A linear relation of the form

$$\alpha \cdot j + \kappa = \frac{1}{A_j} \qquad (18)$$

with constants $\alpha$ and $\kappa$ can be fitted into Fig. 10. The minimum frequency scale is the scale size which exceeds the noise level by $3\sigma$ (see Sect.5) and not necessarily the FWHM of the spike. In both events, the inverse mean peak flux increases with increasing bandwidth scale.

The same comparison has been performed for the individually identified spikes: For every spike, its bandwidth and the peak flux was determined. Subsequently, they were binned into bandwidth bins $[2^{j-1}, 2^j]$ MHz and the mean peak fluxes $A_j$ per bin were determined.

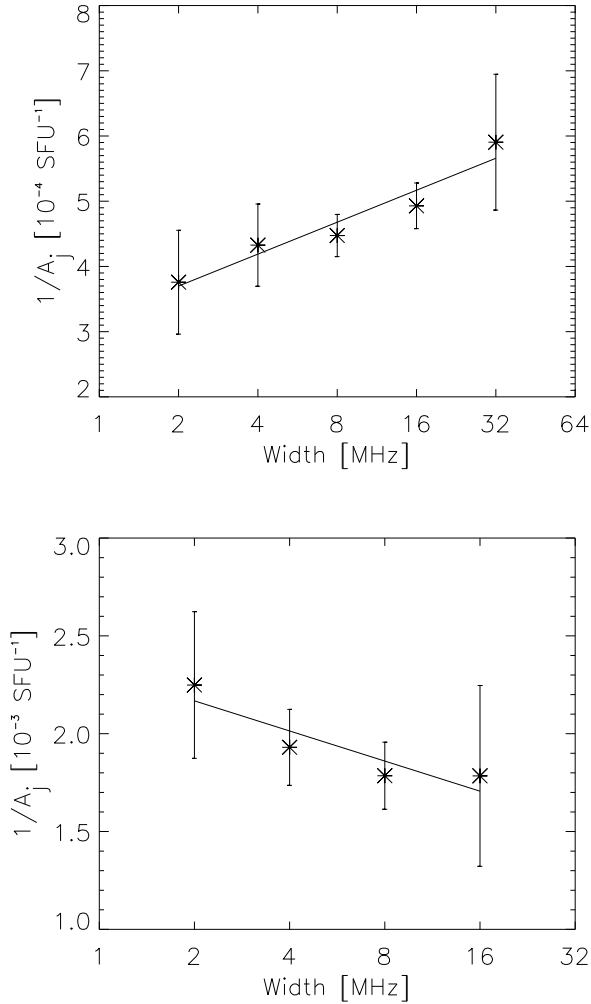

**Fig. 11.** Comparison of bandwidth and inverse mean peak flux $1/A_j$ for both event 1 (top) and event 2 (bottom) for individual spikes, with bandwidth bins $2^j$ MHz, covering bandwidths $[2^{j-1}, 2^j]$ MHz.

Fig. 11 displays the comparison between bandwidth and mean peak flux for both events. Event 1 shows the same relation as determined for the smallest significant scale. On the other hand, event 2 exhibits the inverse tendency with increasing peak flux for increasing bandwidth.

This result shows the difference in the two methods implied: While in both events the peak flux decreases with increasing bandwidth scale, the peak flux may either increase or decrease with increasing FWHM of a fitted Gaussian. The bandwidth scale characterizes *local* properties of spike spectra, whereas the fitted FWHM is a *global* property of each spike.

## 10. Conclusions

The smallest significant bandwidth scales of narrowband radio spikes have been determined by different methods. Scalegram analysis showed structure differing significantly from the mean scalegram down to scales of 2 MHz. Analysis of individual spikes confirmed this, detecting spikes with FWHM down to 1.5 MHz at 900 MHz and 1.2 MHz at 300 MHz respectively.

Based on scalegram analysis of the spectra, the mean significant structure size is 8 MHz at 900 MHz and (marginally) 4 MHz at 300 MHz. Analysis of individual spikes yields a mean FWHM of 7.5 MHz at 900 MHz and 4.8 MHz at 300 MHz.

In relation to the corresponding center frequencies, the minimal FWHMs are 0.17% and 0.41%, respectively. This stringent requirement has to be met by any emission process proposed for spikes. Note that these values are close to the 1 MHz resolution of the instrument. Observations with higher spectral and temporal resolution may even reduce these values.

Eq. 1 can be used to estimate the characteristic velocity $v_{\text{hot}}$ of the masering electrons. From the minimal bandwidths, we derive for the two events $v_{\text{hot}} \lesssim 1.2 \cdot 10^9$ cm/s and $\lesssim 1.9 \cdot 10^9$ cm/s, respectively. The approximate sign applies for a homogenous source. Note that on the other hand, $v_{\text{hot}}$ must exceed the ambient thermal velocity (about $5 \cdot 10^8$ cm/s). Therefore, Eq. 1 and the theory behind it are consistent with the observed minimal bandwidth only if the non-thermal electrons responsible for the emission have low energies and the source is nearly homogenous. The smallest observed bandwidths must be close to the natural width of the emission process.

Mean power spectra of two spike events were computed. Assuming the determined bandwidth scales as limit for relevant frequencies, a power law was fitted to the spectra. Transforming to a scale of characteristic height, and recomputing the power-law exponents, the results are $\log P / \log k = -1.34$, resp. $-1.92$. Both cases do not obviously agree with a Kolmogorov law of $\log P / \log k = -5/3$.

A good correlation between inverse peak flux and spike bandwidth scale or spike FWHM was found. In case of the bandwidth scale, in both cases an decreasing peak flux was found with increasing bandwidth scale. The peak flux decreased in one case and increased in the other with the fitted FWHM bandwidth of individual spikes.

Wavelet filtering and Gaussian fitting has proven to be the superior method for the measurement of the minimum bandwidth. The multiresolution analysis has revealed a different aspect of the bandwidth/amplitude anti-correlation, which however needs to be further investigated.

*Acknowledgements.* Parts of the MRA software is based on an implementation by M. Specht. The observations have been carried out with the help of all members of the Radio Astronomy and Plasma Physics Group. The construction of the Zurich Spectrometers is financially supported by the Swiss National Foundations (grant. no. 20-53664.98).